\documentclass[twoside]{dis04}
\def\runauthor{T.\ Gleisberg et al.}
\def\shorttitle{Monte Carlo models at the LHC}
\begin{document}

\title{Monte Carlo models at the LHC}

\author{T.\ Gleisberg, S.\ H{\"o}che, F.\ Krauss, A.\ Sch{\"a}licke,
        S.\ Schumann, \\ J.\ Winter, G.\ Soff}

\address{Institute for Theoretical Physics\\
TU Dresden, {\bf D}-01062 Dresden, Germany\\
E-mail: krauss@theory.phy.tu-dresden.de}

\maketitle

\abstracts{In this contribution the new event generation framework {\tt SHERPA}
           will be presented, which aims at a full simulation of events at 
           current and future high-energy experiments. Some first results 
           exemplify its capabilities.}

\noindent
Experiments at particle colliders are the most prominent way of testing the basic structure of matter at 
shortest distances and the dynamics underlying the interactions of its fundamental constituents at the 
energy frontier. These tests, however, become increasingly precise; the number of processes as well as the 
complexity of the phenomena involved require an increasingly careful planning of the experimental strategy 
and more and highly sophisticated analysis tools. In order to compare theoretical predictions with actual
experimental results, simulation programs, also known as event generators, have become one of the most
central tools. Over the past decades, these programs have been significantly improved in terms of physics
content and they have grown to highly evolved computer code. In view of the next generation of collider 
experiments it is apparent that also the development of event generators have to keep pace with the new 
great challenges posed by them. Apart from physics issues, reflecting the rising complexity of the 
experiments, transparency and maintenance of these codes start to become an important issue.

\noindent
In order to meet these increasingly demanding requirements, a number of new codes are being constructed 
at the moment. These include the completely new write-ups in {\tt C++} of the well-known Fortran programs 
{\tt Herwig} \cite{Corcella:2000bw}
and {\tt Pythia} \cite{Sjostrand:1993yb}.
Both {\tt Herwig++} \cite{Herwig++} and {\tt Pythia7} \cite{Pythia7} rely on a common event generation
framework called {\tt ThePEG} \cite{ThePEG} and concentrate on the specific implementation of physics
models for different aspects of event simulation. {\tt ThePEG}, in turn, incorporates the organisation and
structure of event generation itself. Another, completely independent approach is represented by the
package {\tt SHERPA} \cite{Gleisberg:2003xi}, also written in {\tt C++}. In the remainder of this 
contribution, this new event generator will be presented in more detail; for details concerning the 
status of the two other {\tt C++}-based event generators, cf.\ the respective presentations by L.\ L{\"o}nnblad
and S.\ Gieseke.

\noindent
One of the construction paradigms of {\tt SHERPA} is the clear separation of specific physics implementation 
and the more abstract rules defining the interplay of the different physics modules. To exemplify this, consider
the case of potentially different codes for the description of hard scattering matrix elements. The {\tt SHERPA}
framework provides an interface called {\tt Matrix\_Element\_Handler} to steer these codes. In turn, this
interface is used by different phases of event generation, like, e.g., the generation of the signal process
or the simulation of the hard underlying event. However, in the following the focus will be on physics issues
only. In its current version, {\tt SHERPA-1.0.4.}, the following physics modules are implemented:
\begin{itemize}
\item Interface to various PDFs: CTEQ \cite{Pumplin:2002vw} and MRST \cite{Martin:1999ww} in their original form
      as well as many other PDFs through the LHAPDF interface in its version 1 \cite{LHAPDFv1}.
\item {\tt AMEGIC++} \cite{Krauss:2001iv} as generator for the matrix elements for hard scattering processes and 
      decays as well as an internal library of analytical expressions for some very constrained set of 
      $2\to 2$ processes. {\tt AMEGIC} contains the full MSSM, where {\tt SHERPA} provides an interface to 
      {\tt Isajet} \cite{Baer:1999sp} for the SUSY particle spectra\footnote{The inclusion of the corresponding 
      Les Houches accord interface \cite{Skands:2003cj} is in preparation.}.
      {\tt AMEGIC++} has exhaustively been tested for a large number of production cross sections for six-body 
      final states at an $e^+e^-$-collider \cite{Gleisberg:2003bi} and various processes at the LHC \cite{MC4LHC}.
\item For multiple QCD bremsstrahlung, i.e.\ the emission of secondary partons, {\tt SHERPA}s own parton shower
      module {\tt APACIC++} \cite{Kuhn:2000dk} is invoked\footnote{In addition to the published version, it has 
      been supplemented by parton showers in the initial state, enabling {\tt SHERPA} to also simulate 
      events with hadronic initial states.}. The merging of the hard matrix elements for multijet production
      and the subsequent parton shower is achieved according to the merging procedure proposed in
      \cite{Catani:2001cc}
      , heavy quarks are treated with 
      corresponding Sudakov form factors \cite{Krauss:2003cr}.
\item Multiple parton interactions, giving rise to the ``hard'' underlying  event, are currently being implemented.
      The corresponding module will be part of the next release of {\tt SHERPA}.
\item Hadronisation of the resulting partons and subsequent hadron decays so far are realized by an interface to
      the corresponding {\tt Pythia} routines. However, a new version of cluster fragmentation \cite{Winter:2003tt} 
      is ready to be fully implemented in the near future.
\end{itemize}

As a first example of the capabilities of the {\tt SHERPA} framework, the production of single electroweak
gauge bosons, i.e. $W$ or $Z$-bosons, is considered in this presentation. The idea underlying the merging
prescription \cite{Catani:2001cc}
is to separate the phase space for parton emission
in two regions, one for jet production, described by appropriate matrix elements, and one for jet evolution,
modelled by the parton shower. The first step is realized by reweighting the matrix elements with Sudakov
form factors and by applying suitable dynamical scales for the strong coupling constant; the second step
translates into vetoing hard parton emission in the subsequent parton shower. A systematic check of this
procedure therefore consists of three steps:
\begin{enumerate}
\item In a first step, the reweighting part is taken as a scale-setting prescription and the thus modified
      matrix elements are compared with suitable higher order calculations. For this, the computer program
      MCFM has been used \cite{Campbell:2002tg}. In Fig.\ \ref{Wexcl} the results for the $p_\perp$ spectra of the 
      first (left panel) and of the first and second jet (right panel) for $Wj$ and $Wjj$ exclusive final states 
      are compared. Both the NLO calculation and the LO results are for ``exclusive'' jets - for the NLO 
      calculation this translates into constraining the phase space of real parton emission, whereas for the 
      reweighted LO results a suitable choice of scale of the Sudakov form factors has to be applied (the
      corresponding minimal $p_\perp$ of the jet).
      In contrast Fig.\ \ref{Wincl} shows results for inclusive final states, 
      i.e., the phase space for the real correction in the NLO calculation is not restricted. For the reweighted LO matrix elements then the choice of scale for the Sudakov form factors is dynamical, namely the actual
      $p_\perp$ of the jet or the softer of the two jets, respectively. This difference also explains the relative
      effect of the higher order correction. The results show an impressive agreement, supporting the
      idea that reweighting the matrix elements at LO takes proper care of higher order corrections. This
      statement, however, has to be taken with a grain of salt: The agreement is for shapes only but not for their total normalisation.
      To have also the normalisation correct, one has to apply a constant $K$-factor given by the ratio of LO
      and NLO total cross section.
      \begin{figure}
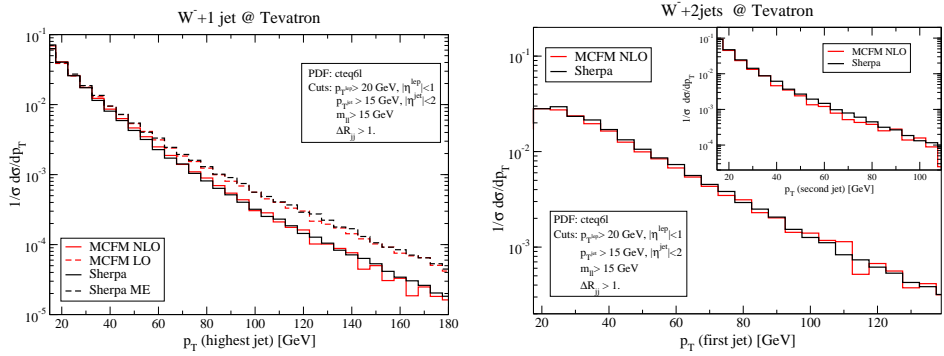

        \begin{center}
          \begin{tabular}{cc}
            \includegraphics[width=6cm]{W1jet_ptjet_excl.eps} &
            \includegraphics[width=6cm]{W2jet_ptjets_excl.eps} 
          \end{tabular}
          \caption{\label{Wexcl}$p_\perp$ spectra of the first (left panel) and of the first and second jet 
            (right panel) of the corresponding NLO calculations for exclusive $Wj$ and $Wjj$ final states and of the 
            reweighted LO matrix elements of {\tt SHERPA} in $p\bar p$ collisions at Tevatron are compared.}
        \end{center}
      \end{figure}
      \begin{figure}
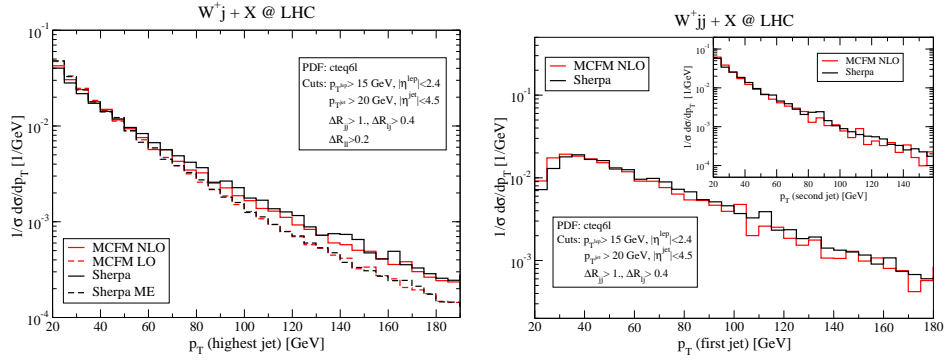

        \begin{center}
          \begin{tabular}{cc}
            \includegraphics[width=6cm]{LHC_Wp1jet_incl.eps} &
            \includegraphics[width=6cm]{LHC_Wp2jet_incl.eps} 
          \end{tabular}
          \caption{\label{Wincl}$p_\perp$ spectra of the first (left panel) and of the first and second jet 
            (right panel) of the corresponding NLO calculations for inclusive $Wj$ and $Wjj$ final states and of the 
            reweighted LO matrix elements of {\tt SHERPA} in $p\bar p$ collisions at Tevatron are compared.}
        \end{center}
      \end{figure}
\item In a next step the way the parton shower is added, including the veto procedure, is controlled. 
      It is important to check that the result for sufficiently inclusive observables is independent on the 
      separation scale between matrix element and parton shower regime\footnote{Of course, if correlations sensitive
      to quantum mechanics are considered, one has to ensure that the matrix elements dominate in the
      relevant region.}; also the independence on the number of extra parton emissions handled by the matrix 
      elements is of relevance. {\tt SHERPA} passes these checks, as can be seen from Figs.\ \ref{Wpt}
      and \ref{Weta}, where the former exhibits the $p_\perp$ spectrum of the $W$-boson at Tevatron, Run II, whereas 
      in the latter its $\eta$ spectrum is depicted, using matrix elements with up to three extra jets. In both figures, 
      different minimal $p_\perp$ between jets or jets and the beam are applied. Also, in all plots there is a 
      second, dashed black line showing the results for a cut of 20 GeV and for up to two extra jets.
      \begin{figure}
        \begin{center}
          \begin{tabular}{cc}
            \includegraphics[width=6cm]{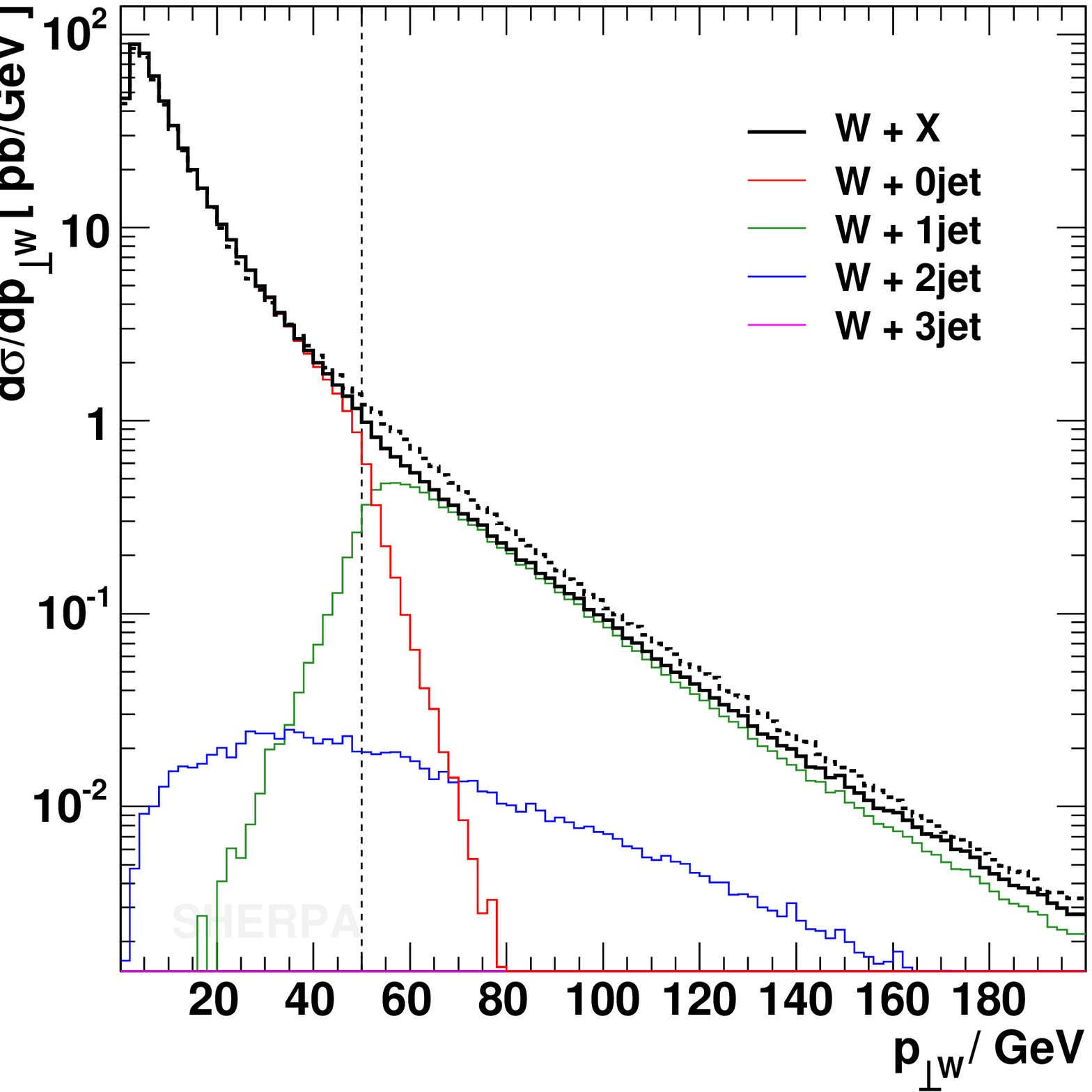}&
            \includegraphics[width=6cm]{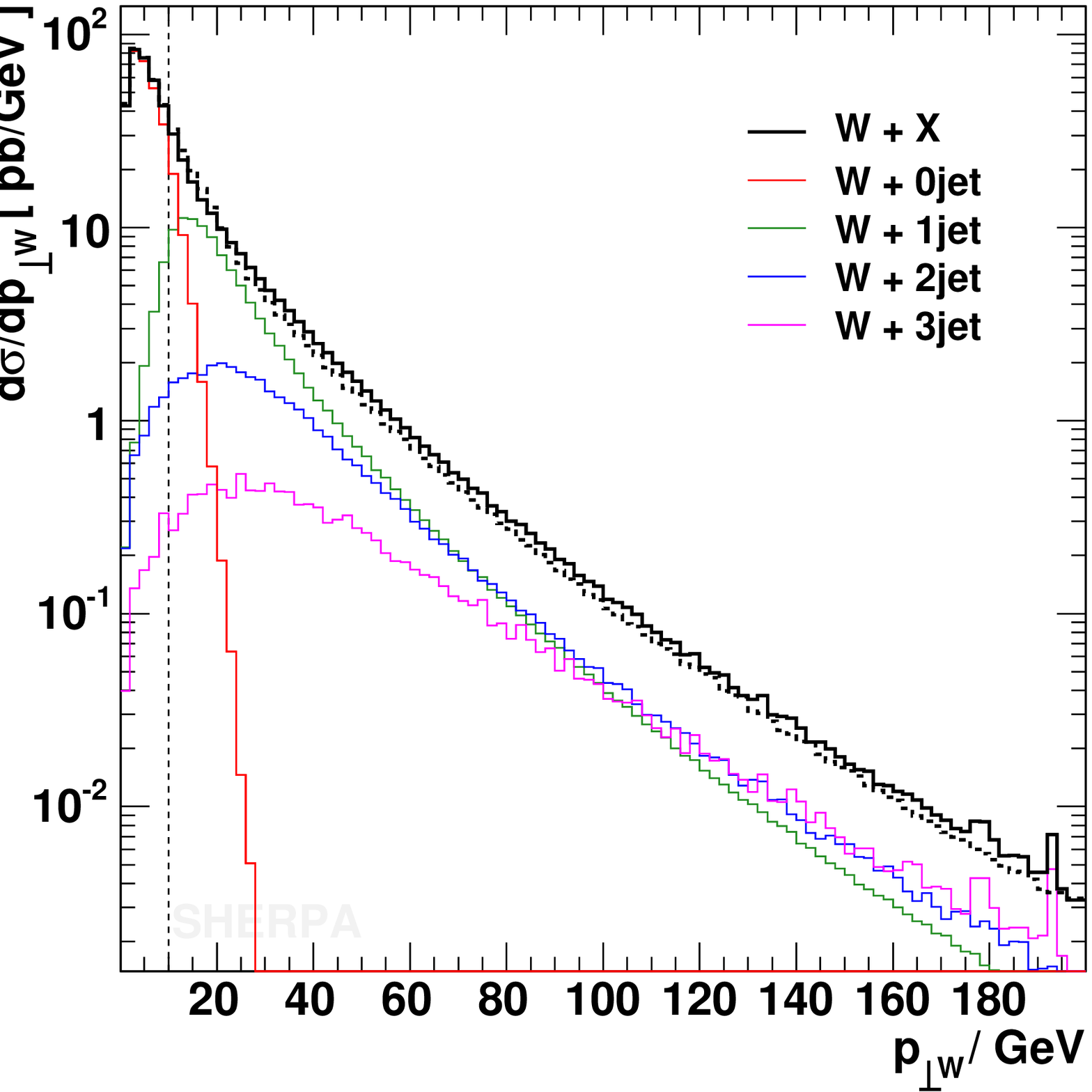}
          \end{tabular}
          \caption{\label{Wpt}$p_\perp$ spectrum of the $W$ at Tevatron, Run II; separation cuts are 50 GeV
          (left) and 10 GeV (right), solid lines are for individual contributions (up to three extra jets)
          and for the total result, the dashed line is for $W$ plus up to two extra jets with a cut of 20 GeV.}
        \end{center}
      \end{figure}
      \begin{figure}
        \begin{center}
          \begin{tabular}{cc}
            \includegraphics[width=6cm]{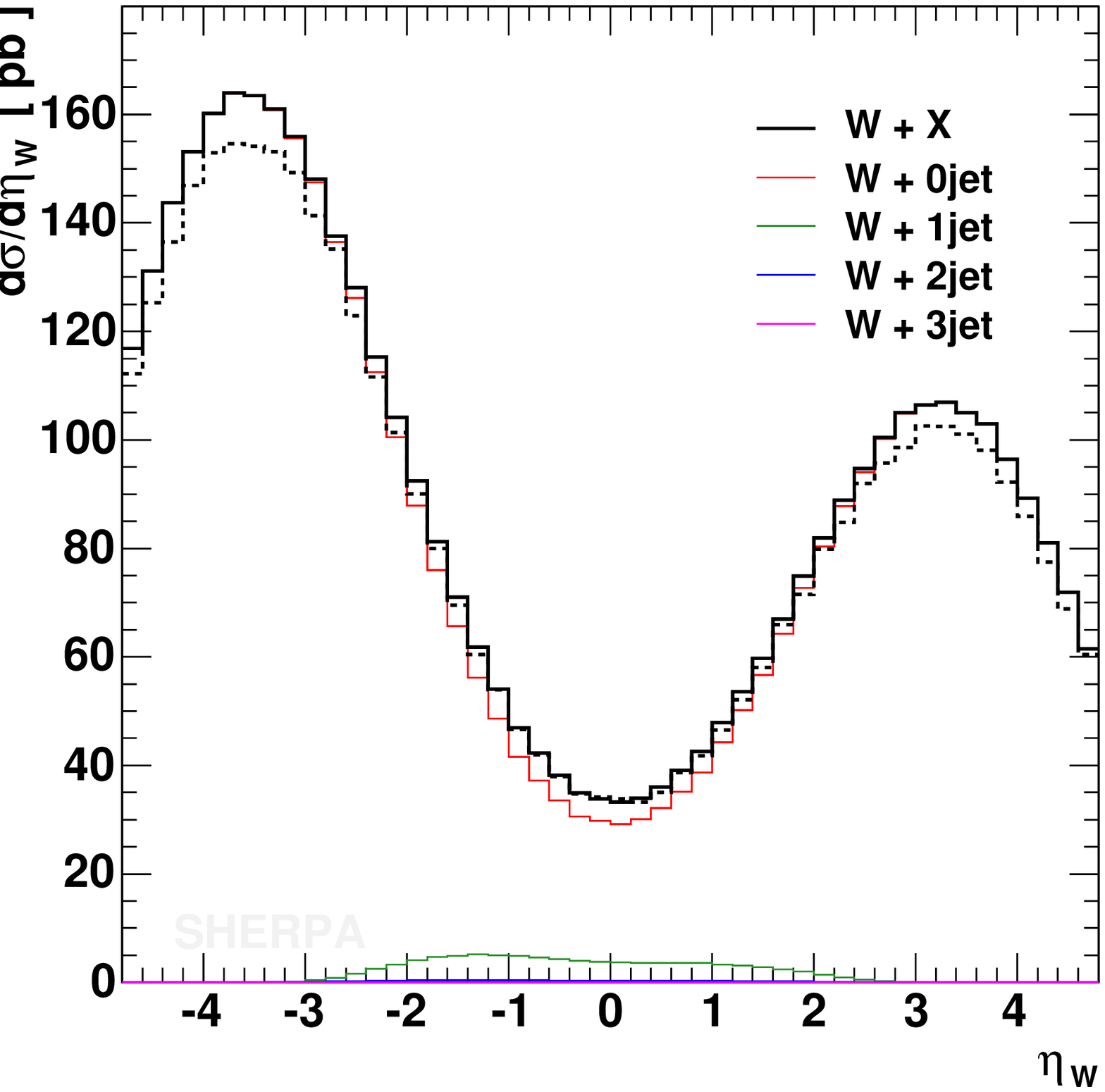}&
            \includegraphics[width=6cm]{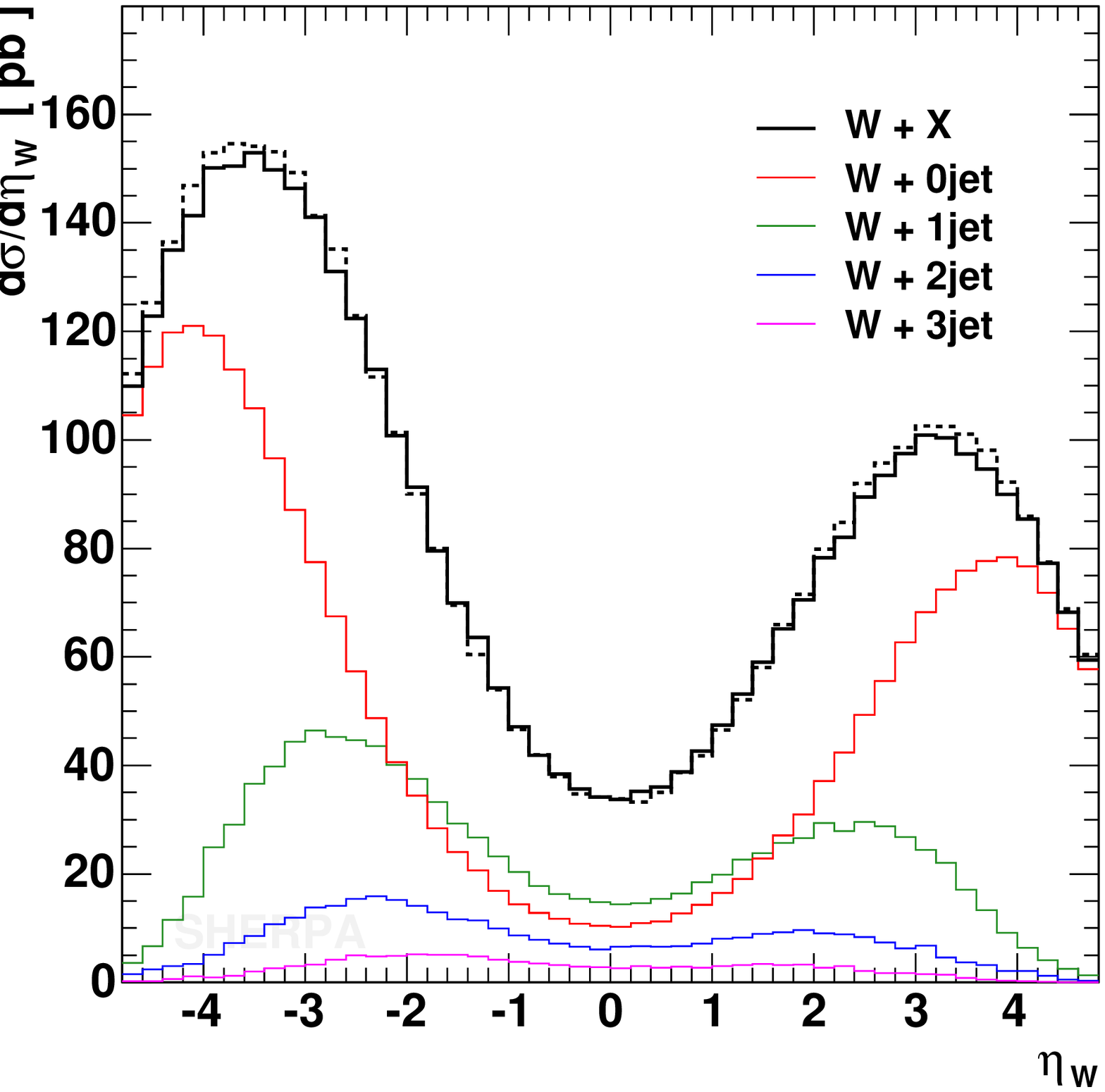}
          \end{tabular}
          \caption{\label{Weta}$\eta$ spectrum of the $W$ at Tevatron, Run II; separation cuts are 50 GeV
          (left) and 10 GeV (right), solid lines are for individual contributions (up to three extra jets)
          and for the total result, the dashed line is for $W$ plus up to two extra jets with a cut of 20 GeV.}
        \end{center}
      \end{figure}
\item Finally the results are compared with experimental data from Tevatron, Run I. In the left panel of
      Fig. \ref{Data} the $p_\perp$ distribution of the $Z$ is displayed \cite{Affolder:1999jh},
      whereas in the right panel the $p_\perp$ distribution of the $W$-boson is shown \cite{Abbott:2000xv}.
      In both cases, {\tt SHERPA}s results are rescaled by a constant $K$-factor.
      \begin{figure}
        \begin{center}
          \begin{tabular}{cc}
            \includegraphics[width=6cm]{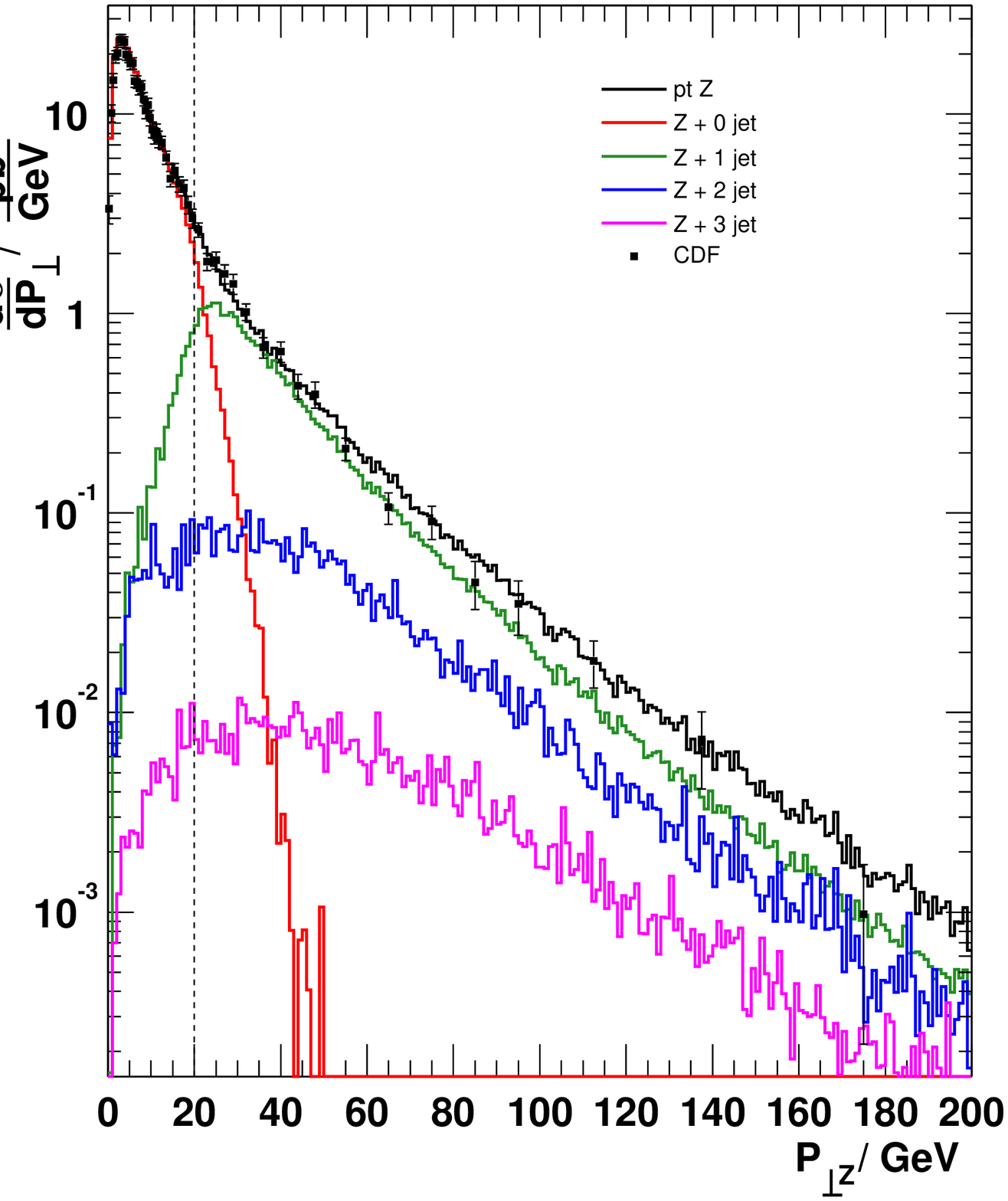} &
            \includegraphics[width=6cm]{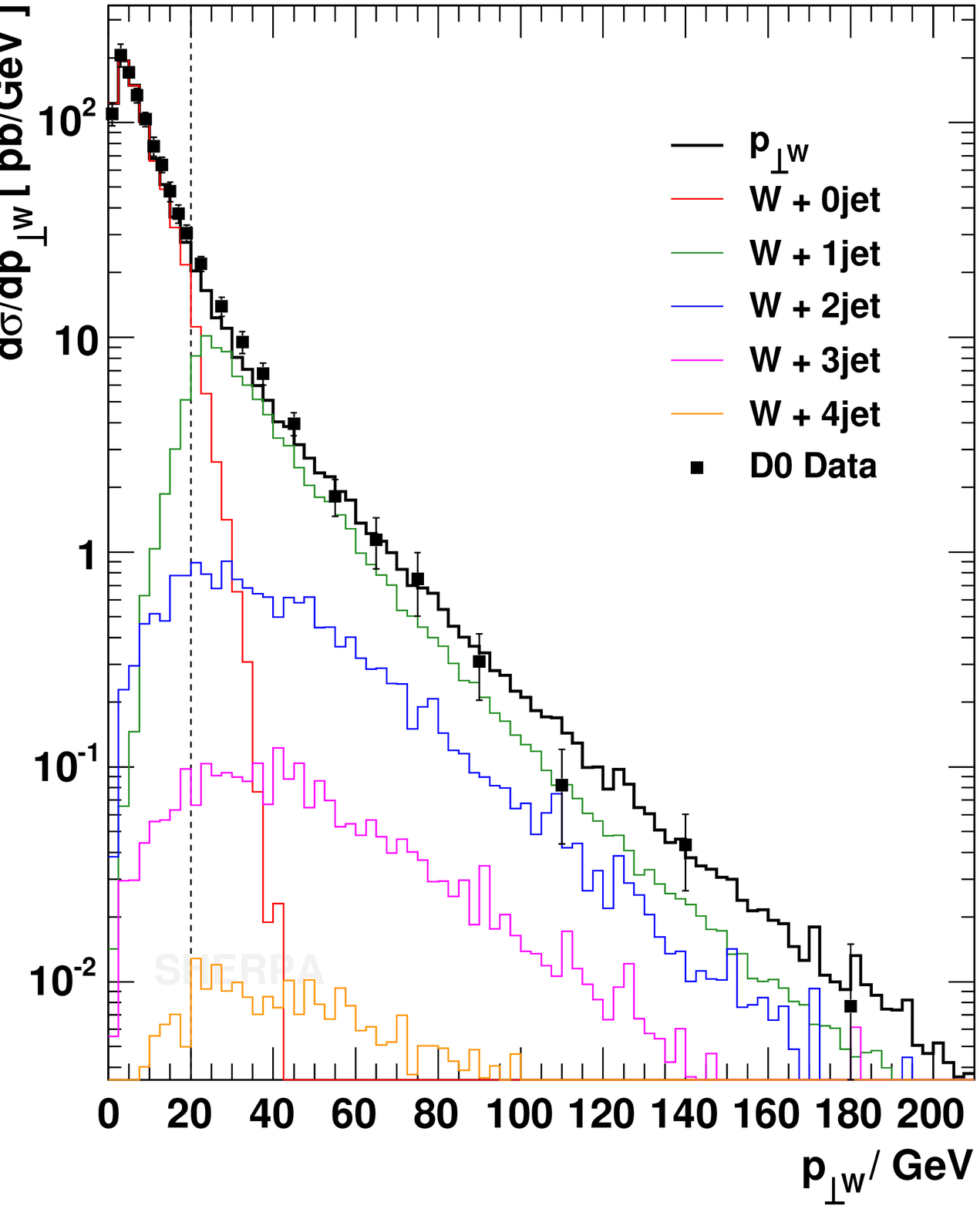}
          \end{tabular}
          \caption{\label{Data}$p_perp$ distribution of the $Z$ (left panel) and of the $W$ (right panel)
                   at Tevatron, Run I, compared with predictions by {\tt SHERPA}, which have been
                   rescaled by a constant $K$-factor.}
        \end{center}
      \end{figure}
\end{enumerate}

\noindent
Taken together, these results prove that the merging as implemented in {\tt SHERPA} is working
in a systematically correct manner; further tests include, e.g., the sensitivity of results to
the choice of scale, the quality in describing more complicated correlations, for instance
of different jets and the simulation of more processes. This programme currently is being worked on,
first preliminary results are very encouraging. This indicates that {\tt SHERPA} is perfectly
suitable to meet the enhanced demands of the community to reliably simulate physics processes
at the next generation of collider experiments.

\section*{Acknolwedgments}
The authors gratefully acknowledge financial support by BMBF, DFG, and GSI. F.K.\ wishes to thank
the organisers of DIS2004 for the extremely pleasant and fruitful atmosphere.

\end{document}